\documentclass[aps,prl,twocolumn,amsmath,amssymbs,superscriptaddress]{revtex4-1}

\usepackage{graphicx}
\usepackage{amsmath,amsfonts,amssymb}
\usepackage{epsfig}
\usepackage{wrapfig}
\usepackage{bbm}
\usepackage[usenames]{color}
\usepackage{array}
\usepackage{times}

\graphicspath{{Figures/}}

\begin{document}

\title{Photonic quadrupole topological phases}

\author{Sunil Mittal}
\affiliation{Joint Quantum Institute, NIST/University of Maryland, College Park, MD 20742, USA}
\affiliation{Department of Electrical and Computer Engineering, and IREAP, University of Maryland, College Park, MD 20742, USA}

\author{Venkata Vikram Orre}
\affiliation{Joint Quantum Institute, NIST/University of Maryland, College Park, MD 20742, USA}
\affiliation{Department of Electrical and Computer Engineering, and IREAP, University of Maryland, College Park, MD 20742, USA}

\author{Guanyu Zhu}
\affiliation{Joint Quantum Institute, NIST/University of Maryland, College Park, MD 20742, USA}
\affiliation{Department of Electrical and Computer Engineering, and IREAP, University of Maryland, College Park, MD 20742, USA}

\author{Maxim~A.~Gorlach}
\affiliation{ITMO University, Saint Petersburg 197101, Russia}

\author{Alexander Poddubny}
\affiliation{Nonlinear Physics Centre, Australian National University, Canberra ACT 2601, Australia}
\affiliation{ITMO University, Saint Petersburg 197101, Russia}
\affiliation{Ioffe Institute, Saint Petersburg 194021, Russia}

\author{Mohammad Hafezi}
\affiliation{Joint Quantum Institute, NIST/University of Maryland, College Park, MD 20742, USA}
\affiliation{Department of Electrical and Computer Engineering, and IREAP, University of Maryland, College Park, MD 20742, USA}
\affiliation{Department of Physics, University of Maryland, College Park, MD 20742, USA}

\begin{abstract}
The topological phases of matter are characterized using the Berry phase, a geometrical phase, associated with the energy-momentum band structure. The quantization of the Berry phase, and the associated wavefunction polarization, manifest themselves as remarkably robust physical observables, such as quantized Hall conductivity and disorder-insensitive photonic transport. Recently, a novel class of topological phases, called higher-order topological phases, were proposed by generalizing the fundamental relationship between the Berry phase and the quantized polarization, from dipole to multipole moments \cite{Benalcazar2017,Benalcazar2017b,Schindler2018,Schindler2018b}. Here, we demonstrate the first photonic realization of the quantized quadrupole topological phase, using silicon photonics. In this 2nd-order topological phase, the quantization of the bulk quadrupole moment in a two-dimensional system manifests as topologically robust corner states. We unambiguously show the presence of localized corner states and establish their robustness against certain defects. Furthermore, we contrast these topological states against topologically-trivial corner states, in a system without bulk quadrupole moment, and observe no robustness. Our photonic platform could enable the development of robust on-chip classical and quantum optical devices with higher-order topological protection.

\end{abstract}

\maketitle


The Berry phase provides a universal framework which relates the robust quantization of physical observables at the boundaries of a non-interacting system to the topological properties of the bulk. For example, in electronic systems, the Berry phase dictates the presence of quantized zero-dimensional (0D) charges in one-dimensional (1D) insulators \cite{Su1979,Zak1989}, quantized charge/spin currents along the 1D-edges of a 2D quantum Hall/spin Hall systems~\cite{Thouless1982,Bernevig2006}, and conducting 2D surfaces in 3D topological insulators~\cite{Liang2011}. In general, this bulk-boundary correspondence relates the quantized charge/currents carried by the $(n-1)$-dimensional boundaries to the quantized electric dipole moments of the $n$-dimensional bulk \cite{King-Smith1993}. Recently, this relationship between the charge moments of the bulk and their boundary manifestations has been generalized from electric dipole to multipole moments, leading to higher-order topological phases \cite{Benalcazar2017,Benalcazar2017b,Schindler2018,Schindler2018b}. For example, a quantized quadrupole moment in a 2D system leads to the presence of quantized charges, which are localized at the 0D corners. This is in contrast to the quantized dipole moment which gives rise to 1D edge currents, in a 2D system.

When formulated in terms of the Berry phase, the bulk-boundary correspondence also applies to neutral bosonic systems, that is, the non-trivial topology of the bulk leads to the emergence of robust boundary states. In fact, topologically robust, localized 0D edge states in 1D systems and propagating 1D edge states in 2D systems have been realized using atomic \cite{Stuhl2015,Atala2013,Aidelsburger2015}, photonic \cite{Wang2009, Hafezi2013, Rechtsman2013, Kraus2012, Cheng2016, Ozawa2018} as well as phononic lattices \cite{Susstrunk2015,Peano2015,Yang2015}. Motivated by the search for higher-order topological phases, the 2D quadrupole topological phases with robust 0D corner states were recently observed using microwaves \cite{Peterson2018,Imhof2018,Serra2018arXiv} and phononic meta-materials \cite{Serra-Garcia2018,Xue2018,Ni2018}. We report on the first optical realization of the quadrupole topological phase, employing the integrated silicon-photonics platform. Using spectroscopic measurements and direct imaging, we reveal the existence of the localized corner states and show that they are robust against certain fabrication disorders which are ubiquitous in nanophotonic systems. Furthermore, by introducing a quadrupole domain boundary in our system, we show that the observed corner states are not artifacts at the physical corners of the lattice. We note that the corner states can also emerge in systems without a quadrupole moment \cite{Noh2018, Li2018}. We study corner states associated with zero bulk quadrupole moment and observe that they are not immune to fabrication disorders.


\begin{figure*}
\centering
\includegraphics[width=0.98\textwidth]{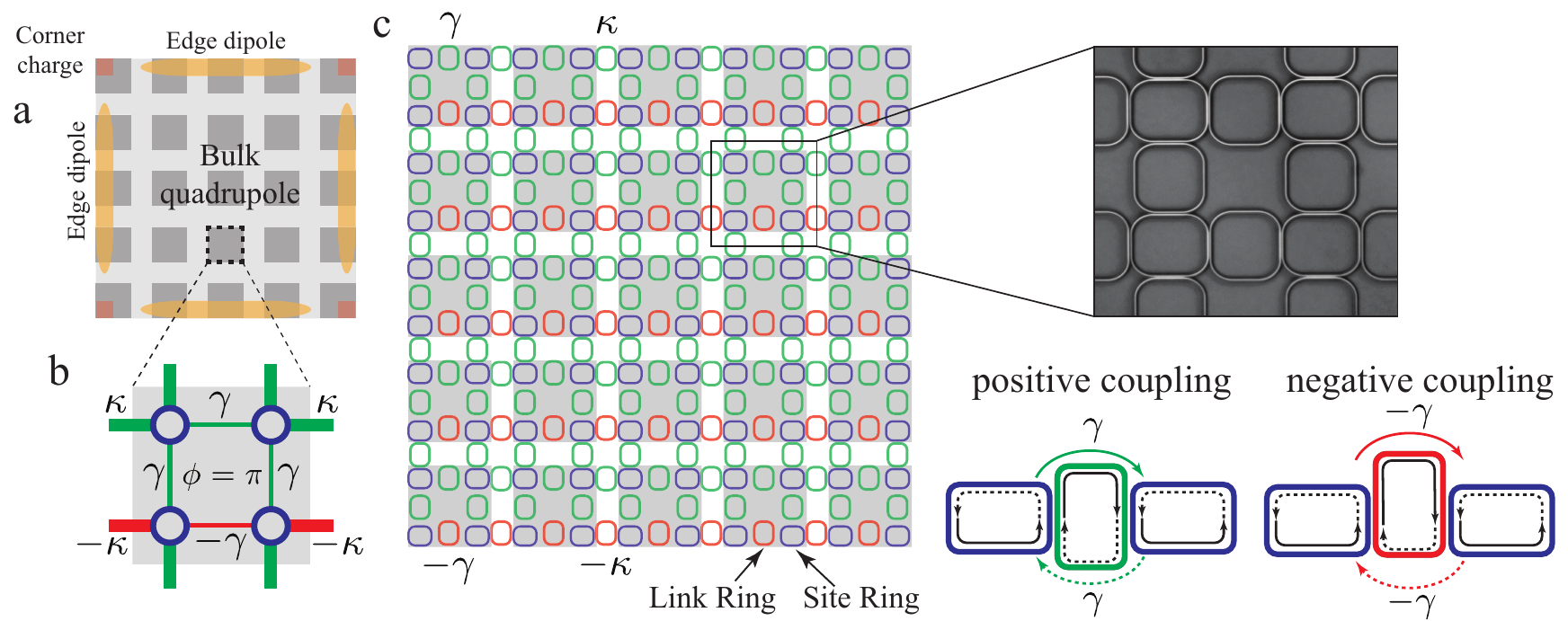}
\caption{
\textbf{a} Schematic of the topological system with quantized quadrupole moment in the bulk, which induces quantized dipole moments on the edges and quantized charges on the corners.
\textbf{b} Schematic of the quadrupole unit cell. Each unit cell consists of four lattice sites with nearest-neighbor couplings, of magnitude $\gamma$. One of the couplings (shown as red) is negative which introduces a gauge flux $\phi = \pi$ per unit cell. The unit-cells are coupled to neighboring cells using coupling strength $\pm \kappa$.
\textbf{c} Schematic of the implementation of the 2D lattice with ring resonators. The unit cell (shaded gray) is composed of four site rings (shaded blue), which are evanescently coupled by four link-rings (shaded green and red). We control the magnitude of inter- and intra-cell couplings by adjusting the gap between the site and the link ring waveguides. To engineer the negative coupling, the position of the link rings (shown in red) are vertically shifted. The link rings (shaded green) corresponding to positive couplings are not shifted. Inset shows a microscope image of the plaquette.
}
\label{fig:1}
\end{figure*}

\begin{figure*}
\centering
\includegraphics[width=0.98\textwidth]{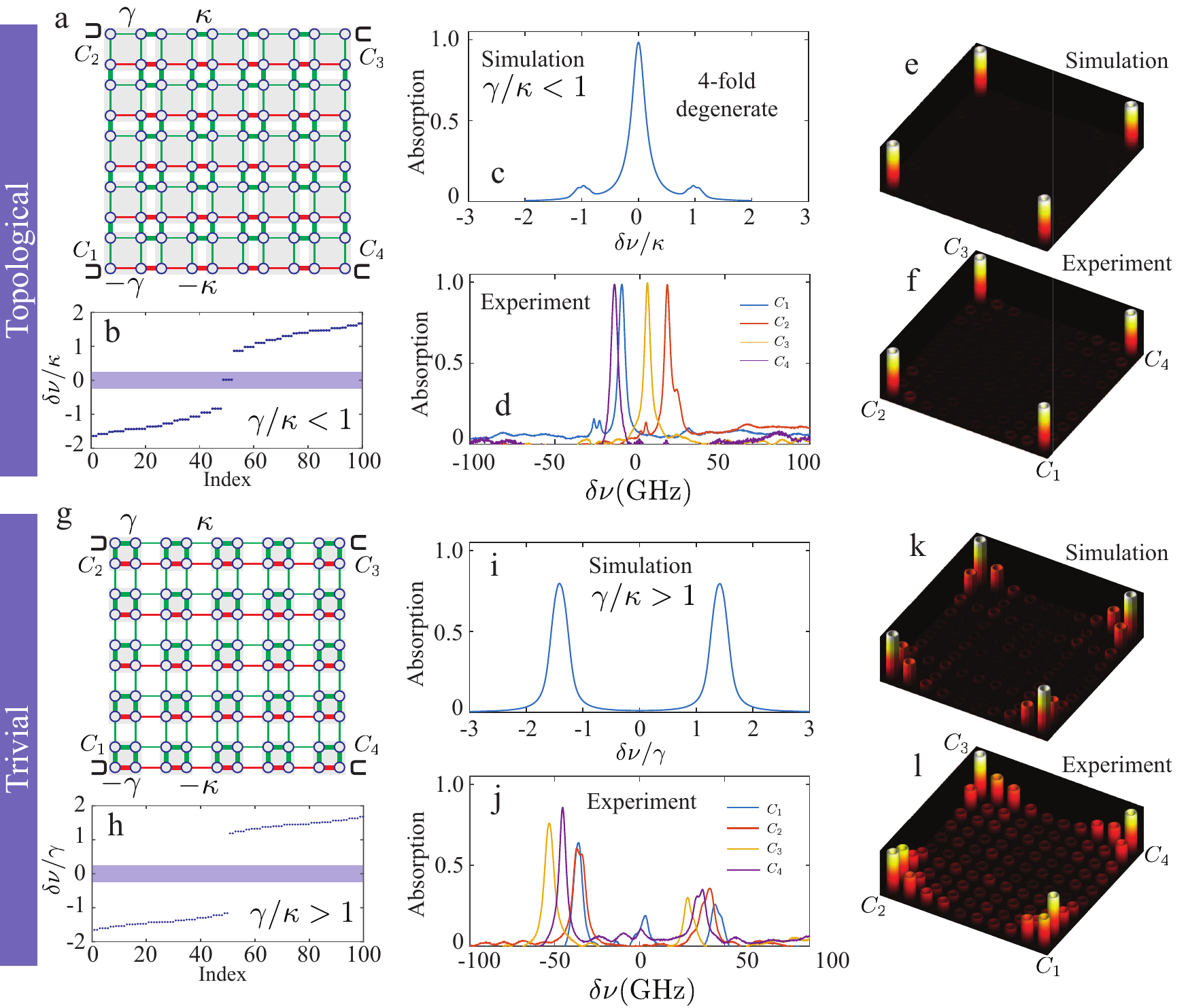}
\caption{
\textbf{a} Schematic of the 2D lattice with quantized quadrupole moment. The circles represent lattice sites. Thick(thin) lines connecting two lattice sites indicate strong(weak) coupling. Green lines indicate positive couplings and red lines indicate negative couplings. The waveguide couplers are located at the four corners.
\textbf{b} Energy spectrum for a lattice with $5 \times 5$ array of unit cells, with $\gamma/\kappa < 1$. The four degenerate corner modes appear in the band-gap.
\textbf{c,d} Simulated and measured power absorption spectra at the corners. In the experimental device, on-site potential disorder breaks the degeneracy of the corner modes.
\textbf{e,f} Simulated and measured spatial intensity profile in the lattice showing localized corner modes. The intensity profiles are summed over excitations from the four corners and correspond to the peak absorption frequencies. Only the intensity at the site rings is shown.
\textbf{g} Schematic of the trivial 2D lattice with $\gamma/\kappa > 1$. \textbf{h} The band-gap is now devoid of any modes. \textbf{i-l} Simulated and measured power absorption spectra and spatial intensity distributions in the absorption bands show absence of localized modes. The intensity profiles are integrated over the two absorption bands and over the four corners.
}
\label{fig:2}
\end{figure*}

Our quadrupole topological system is realized using a two-dimensional (2D) lattice of nanophotonic silicon ring resonators (Fig.\ref{fig:1}, see SI for details on the device fabrication). The unit cell (plaquette) of this lattice consists of four site rings arranged into a square (shaded blue in Fig.\ref{fig:1}). These site rings are evanescently coupled to their nearest neighbors using a set of auxiliary rings, which we call link rings (shaded green and red). We control the magnitude of the coupling strength between the lattice sites by adjusting the gap between the site-ring and the link-ring waveguides. Furthermore, the resonance frequencies of the link rings are detuned from those of the site rings by introducing an extra path length. This allows us to manipulate the sign (phase) of the coupling between the site rings and introduce a synthetic gauge flux threading each square plaquette \cite{Hafezi2011,Hafezi2013}. Specifically, when the link ring coupling two site rings along the x-axis is vertically shifted, the photons hopping towards the right travel a slightly longer path compared to those hopping towards the left (Fig.\ref{fig:1}). This path length difference results in an effective, direction-dependent, hopping phase $\phi$. We choose $\phi = \pi$ such that the shifted link ring (shaded red) introduces a negative coupling. In a unit cell, we arrange the link rings such that $\phi \neq 0$ for only one coupling term, while keeping the amplitude of coupling equal to $\gamma$. This results in a net gauge flux $\pi$ threading the unit cell (Fig.\ref{fig:1}b). To implement a topological system with quantized quadrupole moment \cite{Benalcazar2017}, we realize a 2D array of these unit cells, coupled to their nearest neighbors with coupling strength $\pm \kappa$. Throughout the lattice, we change the sign of the inter-cell coupling $\kappa$, such that a uniform gauge flux $\pi$ penetrates any plaquette, made of $2 \times 2$ array of site rings (see Figs.\ref{fig:1}b,c, and Fig.\ref{fig:2}a). When $\gamma/\kappa < 1$, that is, when the inter-cell coupling is stronger than the intra-cell coupling, the lattice hosts degenerate mid-gap states localized at each of the four corners (Fig.\ref{fig:2}b,e). These states are characterized by topologically invariant integers and protected against certain disorders, due of the presence of spatial inversion symmetry, and the two mirror symmetries (about $x-$ and $y-$ axis) of the lattice which do not commute as a result of the non-zero gauge flux \cite{Benalcazar2017}. In the other case, that is when $\gamma/\kappa > 1$, the lattice is topologically trivial and there are no states in the bandgap (Fig.\ref{fig:2}h).

In our experiment, we fabricate a $5 \times 5 $ array of these unit cells, as shown in Fig.\ref{fig:2}a. We design the system to be in the topological regime with the coupling strengths $\gamma$ and $\kappa$ estimated to be $4.8(2)$ GHz and $29.0(8)$ GHz, such that $\gamma/\kappa \approx 0.17$. To probe the presence of corner states in the lattice, we fabricate a waveguide coupler at each of the corners (Fig.\ref{fig:2}a). This allows us to couple the lattice to a laser and measure the power absorption, as a function of the laser frequency detuning $\left(\delta\nu\right)$ from the ring resonance frequency $\left(\nu\right)$ (see Supplementary Information for further details). Furthermore, we use a microscope objective and an infrared camera to directly image the spatial intensity profile in the lattice, as we sweep the laser frequency. The measured absorption spectra at the four corner sites are shown in Fig.\ref{fig:2}\textbf{d}. We observe very narrow absorption peaks indicating the presence of four isolated mid-gap states in the lattice (Fig.\ref{fig:2}b). Figure \ref{fig:2}f shows the measured spatial intensity profile in the lattice, integrated over the absorption peaks. In agreement with numerical simulation (Fig.\ref{fig:2}e), we see the corner modes are remarkably localized at the corner sites. However, contrary to the theoretical prediction (Fig.\ref{fig:2}\textbf{b,c}), the absorption peaks are shifted with respect to each other, indicating that the corner modes are not degenerate in frequency. This is due the nanofabrication process which introduces very strong disorder in the ring resonance frequencies. The upper bound on the standard deviation in the ring resonance frequencies was measured to be $\approx$ 33 GHz which is very significant compared to the band-gap $2\kappa \approx 58$ GHz. In addition, there is a small disorder in the coupling strengths $\gamma$ and $\kappa$ which was estimated to be $\approx 3 \%$ of the mean values, and also a disorder of $0.1$ (radian) in the gauge flux $\phi$. Remarkably, even in presence of such a strong disorder which is comparable to the band-gap, the corner states are still present and very localized which exhibits their topological protection. We note that the ``zero energy'' in these measurements corresponds to the resonance frequency of a particular longitudinal mode of the ring resonators.

\begin{figure*}
\centering
\includegraphics[width=0.98\textwidth]{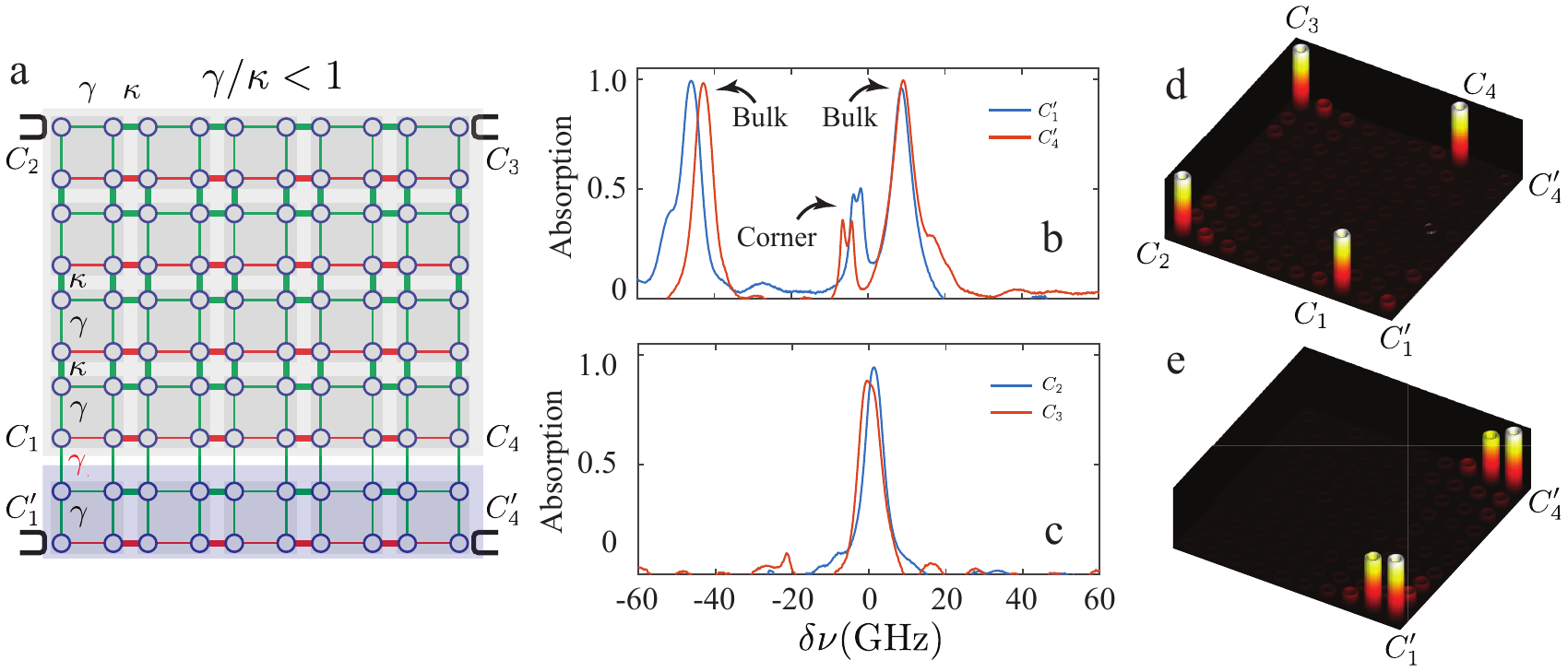}
\caption{
\textbf{a} Schematic of the 2D lattice with a quadrupole domain boundary introduced by breaking the periodicity of the intra-cell hopping $\kappa$. The bottom corners of the quadrupole are labelled as $C1$ and $C4$ and those of the physical lattice as $C'_{1}$ and $C'_{4}$. Absorption measurements are made at the physical corners of the lattice.
\textbf{b,c} Measured absorption spectra at the corners. The middle peak in \textbf{b}, near zero frequency detuning, corresponds to the corner states of the quadrupole system.
\textbf{d} Measured spatial intensity distribution showing localized corner states. The intensity distribution is summed over the single absorption peak for $C_{2}$ and $C_{3}$ and only over the middle peak for $C'_{1}$ and $C'_{4}$.
\textbf{e} Measured intensity distribution corresponding to the two sidebands, away from zero frequency detuning, of the absorption spectra in \textbf{b}. These are bulk modes of the non-quadrupole domain of the lattice.
}
\label{fig:3}
\end{figure*}

\begin{figure*}
\centering
\includegraphics[width=0.98\textwidth]{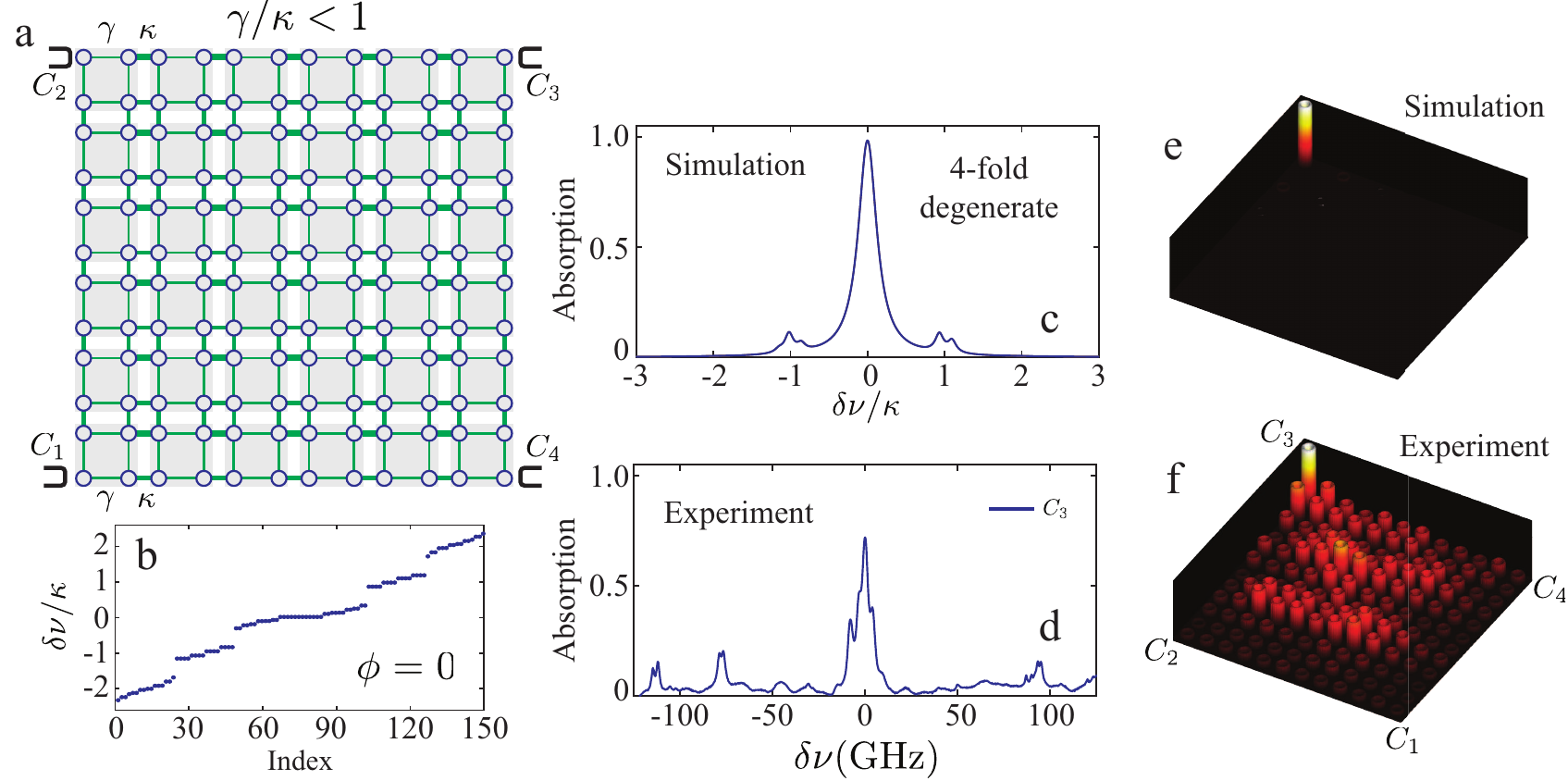}
\caption{
\textbf{a} Schematic of the 2D lattice with $\gamma/\kappa < 1$, but all positive couplings, that is, zero gauge flux. We fabricated a $6 \times 6$ array of unit cells for this device.
\textbf{b} The energy spectrum shows four zero-energy modes with wavefunction localized at the corners. But there is no bandgap at zero energy.
\textbf{c} Simulated absorption spectrum at the corner, for a pure system (no disorder), shows predominantly single absorption peak, similar to that of the quadrupole system.
\textbf{d} Corresponding intensity distribution for excitation at corner $C_{3}$ also shows a single mode localized at the corner. \textbf{e,f} Experimentally measured absorption at corner $C_{3}$ and the corresponding intensity distribution indicate significant coupling of corner modes to the bulk modes because they are not robust against fabrication disorder (additional data in S.I.). The intensity distribution is integrated over the absorption band.
}
\label{fig:4}
\end{figure*}

To show the absence of corner states in a trivial system, we analyze the scenario where the inter-cell hopping is much stronger than the intra-cell hopping, that is, $\gamma/\kappa > 1$ (Fig.\ref{fig:2}g). For this system, we swap the two coupling strengths, that is, now $\gamma \approx 29$ GHz and $\kappa \approx 4.8$ GHz, which gives $\gamma/\kappa \approx 5.8$. As such, the lattice is topologically trivial and the mid-gap modes are absent (Fig.\ref{fig:2}h). Fig.\ref{fig:2}j shows the measured power absorption spectrum at the four corners and Fig.\ref{fig:2}l shows the spatial intensity distribution integrated over the absorption bands. In contrast to Fig.\ref{fig:2}d, the spectra in Fig.\ref{fig:2}j show two broad absorption bands centered around $\pm \sqrt{2}\gamma$, corresponding to the bulk modes of the lattice \cite{Benalcazar2017}. Furthermore, the observed spatial intensity distributions (Fig.\ref{fig:2}l) confirm that these absorption bands do not correspond to localized modes and are smeared into the bulk of the lattice.

Next, we demonstrate that the observed corner states are not artifacts due to some defects at the physical corners of the lattice. We break the periodicity of the intra-cell coupling strength such that it creates a domain boundary, with the unperturbed quadrupole domain, this time consisting of an array of $5 \times 4$ unit cells (Fig.\ref{fig:3}a). In this configuration, the corners $C_{2}$ and $C_{3}$ of the quadrupole domain coincide with the physical corners of the lattice whereas, corners $C_{1}$ and $C_{4}$ of the quadrupole domain are shifted from the physical corners, labeled $C'_{1}$ and $C'_{4}$ in Fig.\ref{fig:3}a. As before, the waveguide couplers for absorption measurements are located at the physical corners of the lattice. Fig.\ref{fig:3}b,c show the measured absorption spectra. The spectra at corners $C_{2}$ and $C_{3}$ show a single absorption peak and the associated spatial intensity distribution (Fig.\ref{fig:3}d) shows localized corner states at $C_{2}$ and $C_{3}$, consistent with our previous observation. However, the spectra measured at corners $C'_{1}$ and $C'_{4}$ show three absorption peaks. The spatial intensity distribution corresponding to the middle absorption peak around zero frequency detuning, in fact, shows excitation of localized modes at the corners $C_{1}$ and $C_{4}$ of the quadrupole domain (Fig.\ref{fig:3}d). In contrast, the intensity distribution in the other two sidebands, away from zero detuning, occupies the lattice sites outside the quadrupole domain (Fig.\ref{fig:3}e). The power absorption in the middle band is lower relative to the sidebands, mainly because the corner states of the quadrupole are very weakly coupled to the physical corner.

To further explore the significance of topological protection of the observed corner modes in the quadrupole system, we compare them with those emerging in a system with zero quadrupole moment. We fabricated a very similar lattice with zero gauge flux, that is, where all the couplings are positive (Fig.\ref{fig:4}a). The inter-cell and the intra-cell couplings in this lattice still correspond to the topological case (Fig.\ref{fig:2}a), that is, $\gamma/\kappa < 1$. Because of the absence of the gauge flux, the two mirror symmetries (about the $x-$ and the $y-$axis) commute and the net quadrupole moment is zero (see S.I.). Numerical simulation results, in the absence of disorder, show that this system hosts zero-energy modes localized at the corners, similar to those of the quadrupole system (Fig.\ref{fig:4}b,c,e). However, unlike the quadrupole system, the measured absorption spectrum at the corner shows a multi-mode structure indicating that the corner states are not completely isolated from the bulk states (Fig.\ref{fig:4}d,f, additional data in the S.I.). The measured spatial intensity distribution confirms that these corner modes indeed couple to the bulk modes of the lattice. This observation suggests that these corner modes, in the zero gauge flux lattice, are susceptible to the fabrication disorder and are not topologically protected. This lack of robustness against disorder is also evident in the energy spectrum of the device, where there is no band-gap at zero energy. Consequently, an on-site potential disorder, which in our experimental setup is the most dominant source of disorder, can easily couple these corner modes to the bulk modes located near zero energy (see supplementary information for more details).


In this work, we have demonstrated an integrated photonic quadrupole topological phase supporting robust corner states immune to disorder in coupling strengths and on-site potential. Our versatile integrated photonics platform opens the route to explore even richer physics in the context of higher-order topological phases, for example, those associated with different choices of gauge flux in the lattice or with asymmetric coupling strengths. Moreover, by integrating active features, e.g., using thermal heaters or electro-optic modulators \cite{Mittal2016}, our scheme could allow selective flux insertion or manipulation of the corner or the edge modes of the system and hence, the realization of adiabatic charge pumping in high-order topological systems {\cite{Benalcazar2017b}}. Furthermore, the integration of gain/loss in this photonics realization could enable the exploration of active optical devices, such as topological lasers \cite{St-Jean2017, Bahari2017, Bandres2018}, quantum light sources \cite{Mittal2018}, and non-Hermitian physics, in the context of higher-order topological physics.

Note: During the submission of this manuscript, we became aware of other related works \cite{Ota2018, ElHassan2018} which realize corner states of light using generalizations of the SSH model. However, these models lack gauge fields and are not direct realizations of the quadrupole topological phases.

\textbf{Acknowledgements:} This research was supported by AFOSR-MURI FA9550-14-1-0267, Sloan Foundation and the Physics Frontier Center at the Joint Quantum Institute. AP and MG have  been supported by the Russian Foundation for Basic Research Grant No. 18-29-20037 and the Foundation for the Advancement of Theoretical Physics and Mathematics BASIS.  AP also acknowledges partial support by the Australian Research Council. We thank Hossein Dehghani, Yuri Kivshar and Daniel Leykam for discussions, Evan Yamaguchi and Jonathan Vannucci for experimental help.

\bibliographystyle{NatureMag}
\bibliography{Quad_Topo_Biblio}

\end{document}


\title{Supplementary Information: Photonic quadrupole topological phases}

\author{Sunil Mittal}
\affiliation{Joint Quantum Institute, NIST/University of Maryland, College Park, MD 20742, USA}
\affiliation{Department of Electrical and Computer Engineering, and IREAP, University of Maryland, College Park, MD 20742, USA}

\author{Venkata Vikram Orre}
\affiliation{Joint Quantum Institute, NIST/University of Maryland, College Park, MD 20742, USA}
\affiliation{Department of Electrical and Computer Engineering, and IREAP, University of Maryland, College Park, MD 20742, USA}

\author{Guanyu Zhu}
\affiliation{Joint Quantum Institute, NIST/University of Maryland, College Park, MD 20742, USA}
\affiliation{Department of Electrical and Computer Engineering, and IREAP, University of Maryland, College Park, MD 20742, USA}

\author{Maxim~A.~Gorlach}
\affiliation{ITMO University, Saint Petersburg 197101, Russia}

\author{Alexander Poddubny}
\affiliation{ITMO University, Saint Petersburg 197101, Russia}
\affiliation{Ioffe Institute, Saint Petersburg 194021, Russia}

\author{Mohammad Hafezi}
\affiliation{Joint Quantum Institute, NIST/University of Maryland, College Park, MD 20742, USA}
\affiliation{Department of Electrical and Computer Engineering, and IREAP, University of Maryland, College Park, MD 20742, USA}
\affiliation{Department of Physics, University of Maryland, College Park, MD 20742, USA}

\maketitle

\section{Experimental Setup}

Our topological devices are realized using the silicon-on-insulator (SOI) platform. The ring resonator waveguides are made of silicon and are buried in a layer of silicon oxide. The waveguide cross-section was designed to be 510 nm in width and 220nm in height so that it supports only a single transverse-electric (TE) mode at telecom wavelengths. The devices were fabricated in a standard CMOS fabrication facility, at IMEC Belgium.

To measure the power absorption spectra of the devices, we used a fiber-coupled, frequency tunable, continuous-wave laser (CW). The laser output was coupled to the devices using grating couplers, with a coupling efficiency of $\approx 6$ dB per coupler. The unabsorbed light was routed to another grating coupler, collected in a fiber and measured using a photodetector. The measured absorption spectra include the actual power absorbed by the device, the losses introduced by the two grating couplers and also the fiber connectors. These additional losses can be easily measured at frequencies far detuned from the ring resonances where power absorption by the resonators is negligible. The absorption spectra reported in the main text have been corrected for these losses.

\linespread{1.0}
\begin{figure}[h]
\centering
\includegraphics[width=0.78\textwidth]{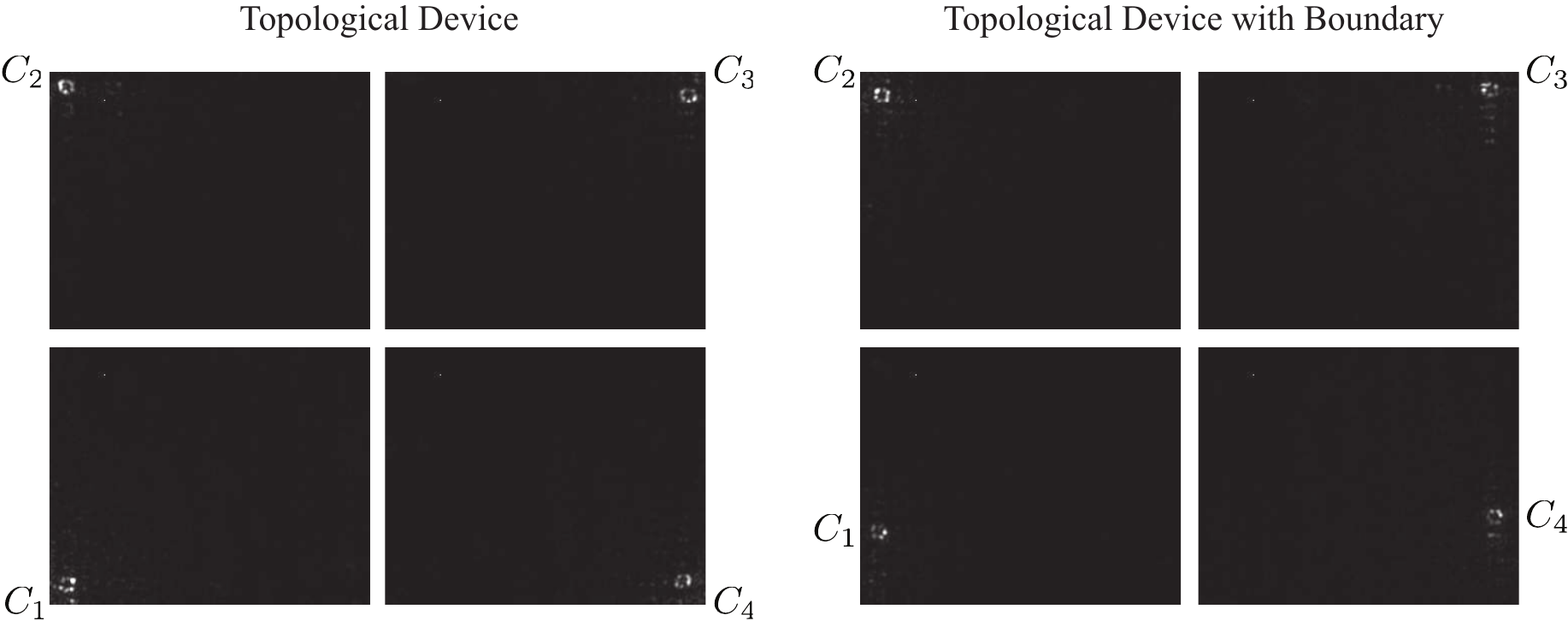}
\caption{Raw camera images showing corner states for the topological devices reported in Fig.2\textbf{d} and Fig.3\textbf{d} of the main text.}
\label{fig:S1}
\end{figure}
\linespread{2.0}

To image the spatial intensity profiles in the lattice, we used a microscope objective (10x) to collect the light scattered from the surface roughness of the ring resonator waveguides. The light was subsequently imaged on a high-sensitivity InGaAs camera ($320 \times 256$ pixels) using a variable zoom lens system. The raw images showing the corner states, corresponding to Fig.2\textbf{d} and Fig.3\textbf{b} of the main text, are shown in Fig.\ref{fig:S1}. Note that the camera positions were changed while imaging different corners to suppress scattering from input fiber.

\section{Verification of $\pi$ gauge flux through a unit-cell}

\linespread{1.0}
\begin{figure}
\centering
\includegraphics[width=0.78\textwidth]{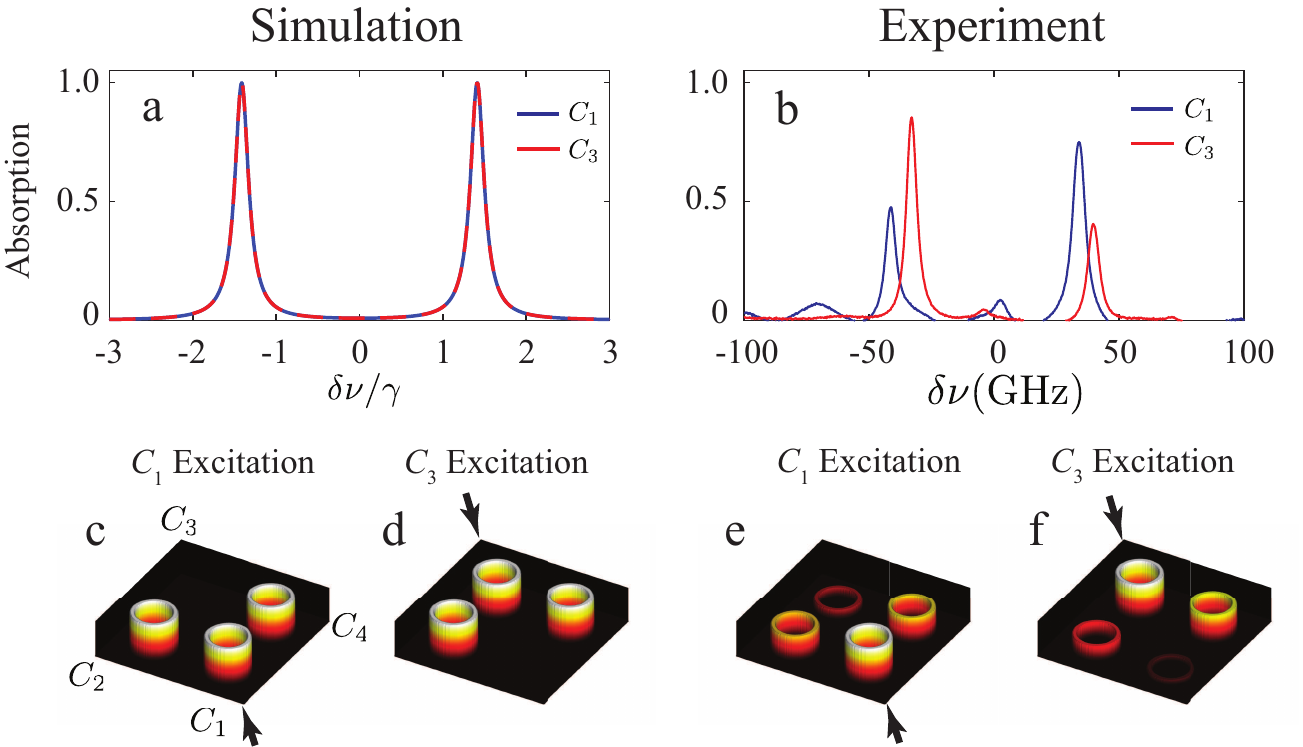}
\caption{
\textbf{a,b} Simulated and measured absorption spectrum of the unit-cell, at the two diagonally opposite corners $C_{1}$ and $C_{3}$. Each spectra shows only two absorption peaks, consistent with doubly-degenerate energy eigenvalues of a quadrupole \cite{Benalcazar2017,Peterson2018}. \textbf{c-f} The corresponding spatial intensity profiles shows absence of intensity at corner $C_{3}$ when $C_{1}$ is excited and vice-versa. The intensity profiles for each excitation is integrated over both the absorption peaks.
}
\label{fig:S2}
\end{figure}
\linespread{2.0}

To verify that the synthetic gauge flux threading the plaquette is indeed $\pi$. When $\phi = \pi$, we fabricate a single unit-cell of the quadrupole lattice. When the flux $\phi = \pi$, this unit cell exhibits two pairs of degenerate eigenvalues at $\pm \sqrt(2) \gamma$, where $\gamma$ is the coupling strength between the site rings \cite{Benalcazar2017}. These eigenvalues can be probed using absorption spectroscopy (Fig.\ref{fig:S2}~\textbf{a}). The measured results for absorption at corners $C_{1}$ and $C_{3}$ are shown in Fig.\ref{fig:S2}~\textbf{b}. The presence of only two absorption peaks in the spectra confirms that the gauge flux $\phi = \pi$. The observed asymmetry in the two absorption peaks is because of the disorder-induced frequency mismatch between the ring resonance frequencies. Furthermore, using direct imaging of the spatial intensity profiles (Fig.\ref{fig:S2}~\textbf{e,f}), we observe that an excitation at corner $C_{1}$ results in negligible intensity (probability density) at the diagonally opposite corner $C_{3}$ and vice-versa. This is because of the destructive interference introduced by the gauge flux $\pi$. For this single unit-cell device, the coupling strength between the site rings was estimated to be $\approx 26$ GHz.

\section{Characterization of the disorder}

The nano-fabrication process introduces various disorders into our devices. The most significant disorder affecting our device is the on-site potential, that is, the mismatch between the site ring resonance frequencies. In addition, there is a disorder in the coupling strengths and also in the gauge flux $\phi$, which is essentially a manifestation of the mismatch between link ring resonance frequencies. We used add-drop filters (ADFs) to characterize these disorders and also the coupling strengths $\gamma$ and $\kappa$. The dimensions (length, bend radius) of the ring resonators used in ADFs matched exactly the dimensions of those used in our 2D devices. But, to allow for independent measurements of the coupling strengths, the ADFs had different coupling gaps between the rings and the input-output waveguides. The through and the drop port spectra were measured for five different ADFs (with different coupling strengths) fabricated on each chip. The standard deviation of the resonance frequencies was calculated for each chip and then averaged over ten chips (total fifty ADFs), to give $U \approx 32.8$ GHz. The measured contrast and the bandwidth of the spectra yielded $\gamma = 4.8$ GHz and $\kappa = 29.0$ GHz as the mean values. The standard deviation in the coupling strengths was $\approx 4 \%$. The standard deviation in the gauge flux $\phi$ was calculated to be $\approx 0.1$ using the estimated value of $U$. Also, the loss in the resonators was estimated to be $\approx 1.8$ GHz, with a standard deviation of $\approx 20\%$.

\section{Edge Excitation for Quadrupole Device}

\linespread{1.0}
\begin{figure}
\centering
\includegraphics[width=0.98\textwidth]{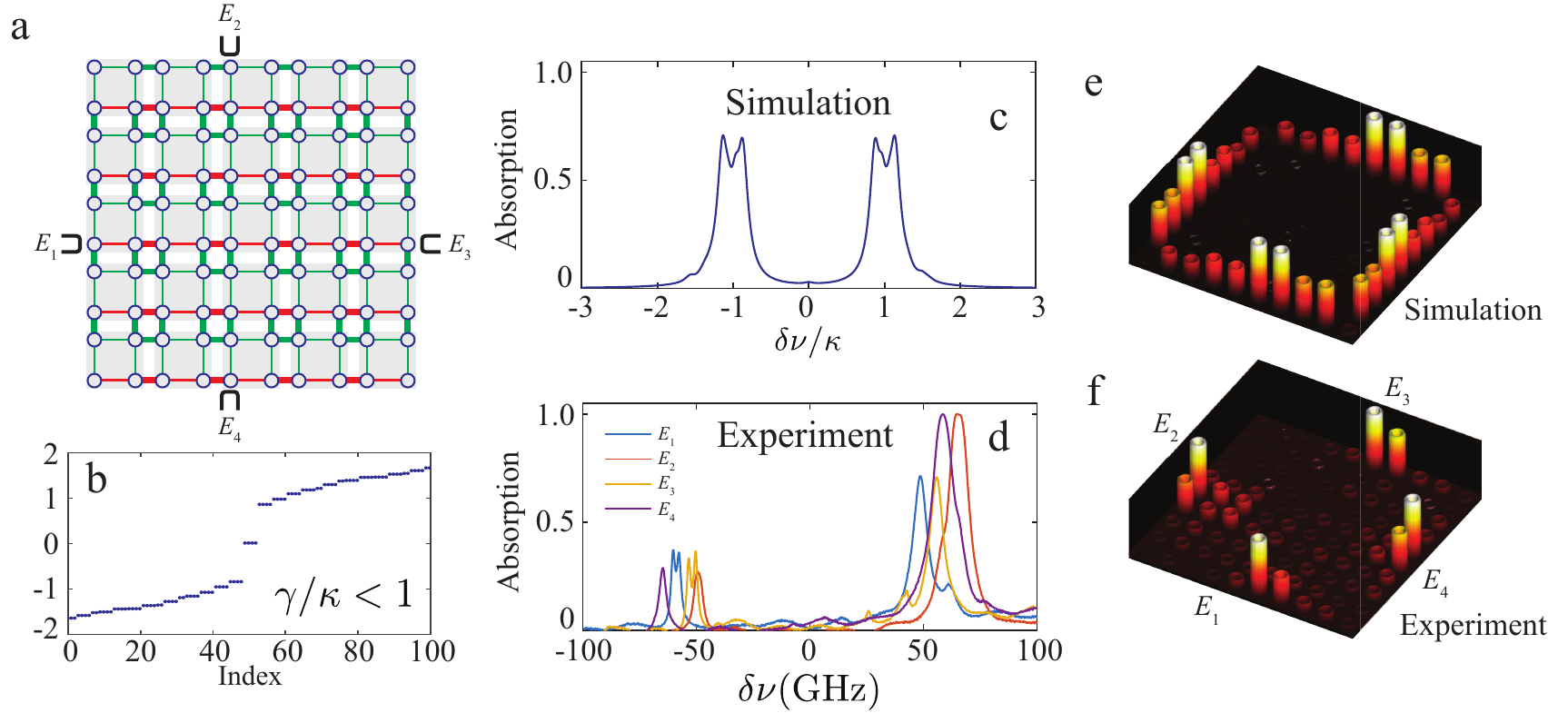}
\caption{\textbf{a} Schematic of the quadrupole device $\left(\frac{\gamma}{\kappa} < 1, \phi = 0 \right)$, showing coupler positions for the edge excitations. \textbf{b} Simulated eigenvalue spectra. \textbf{c,d} Simulated and measured power absorption spectra at the four edges of the device, respectively. The simulated spectra, for a device with no disorder, are four-fold degenerate. \textbf{e-f} Corresponding spatial intensity profiles. }
\label{fig:S3}
\end{figure}
\linespread{2.0}

In addition to the corner modes, the quadrupole topological system also exhibits localized edge states near $\delta\nu = \pm kappa$ \cite{Benalcazar2017,Benalcazar2017b}. The edge states can be probed using waveguide couplers at the edges of the lattice (Fig.\ref{fig:S3}). Fig.\ref{fig:S3} shows simulated and measured power absorption spectra at the edges and the corresponding spatial intensity profiles in the lattice. In contrast to simulation, we observe two broad absorption peaks near the expected bulk band positions $\left(\delta\nu = \pm \sqrt{2} kappa\right)$, indicating a coupling between the edge and the bulk states. This is because there is no bandgap between the edge and the bulk bands and therefore, the on-site potential disorder in the experimental system can couple the two bands.

\section{2D Lattice with zero gauge flux}

\linespread{1.0}
\begin{figure}[h]
\centering
\includegraphics[width=0.78\textwidth]{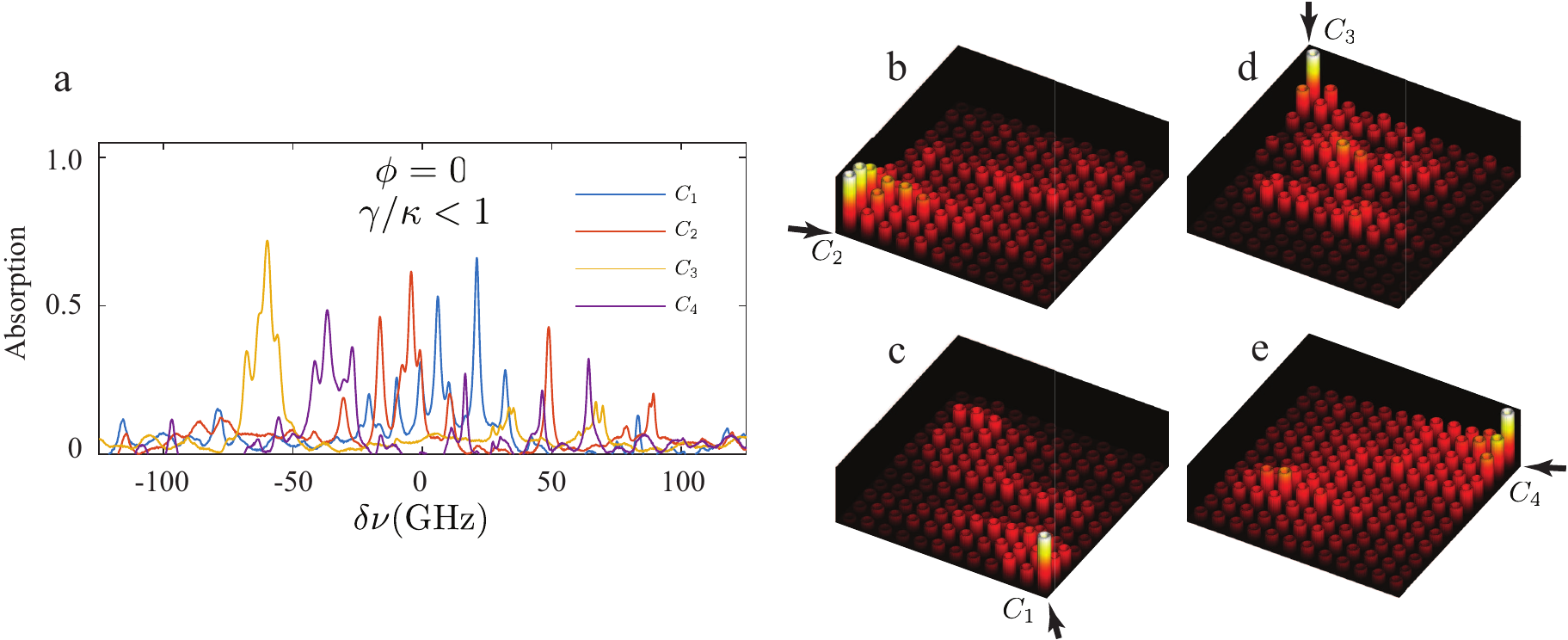}
\caption{\textbf{a} Measured power absorption spectra at the four corners of the device with all positive couplings, that is, $\phi = 0$. \textbf{b-e} Corresponding spatial intensity profiles showing coupling of corner and bulk modes. The intensity profiles are integrated over the absorption peaks.}
\label{fig:S4}
\end{figure}
\linespread{2.0}

To highlight the robustness of the topological corner states in a quadrupole system, we compare them against those emerging in a system with no gauge flux $\left(\phi = 0 \right)$ which results in zero quadrupole moment. Numerical simulation results for such a system $\left(\phi = 0 \right)$ show sharply localized corner states in the absence of disorder (Fig.4\textbf{c,e} of the main text). However, the measured absorption spectra (Fig.\ref{fig:S4}) show multi-mode behavior and are consistent with the measured intensity profiles which reveal significant coupling of corner modes to the bulk modes. This clearly demonstrates that the corner states in a system with zero quadrupole moment are not robust against disorder, unlike those associated with quadrupole topological phases.

\section{Topological transition realized by varying the gauge flux in the lattice}

The proposed integrated silicon photonic platform allows one to realize even richer variety of topological systems by tuning the phase of hopping amplitudes. As an interesting illustration, we examine here the transition between the  two-dimensional tight-binding system with all positive couplings with ungapped spectrum and the standard quadrupole insulator characterized by the gapped bands as the magnitude of flux through each plaquette of the lattice is varied from 0 to $\pi$.

\begin{figure}[h]
\centering
\includegraphics[width=0.98\textwidth]{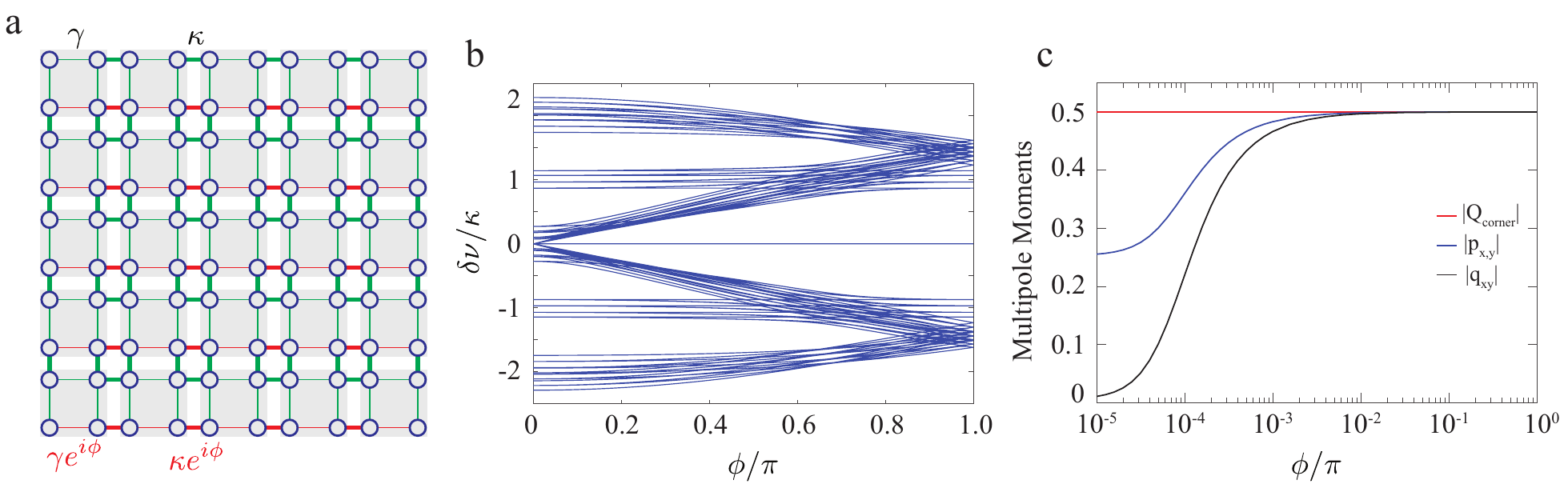}
\caption{
\textbf{a} The schematic of tight-binding system under study realized as an array of evanescently coupled ring resonators. Coupling is ensured via the auxiliary rings.
\textbf{b} Numerically computed evolution of the system spectrum as the gauge flux is varied from 0 to $\pi$. Opening of a topological gap is observed and the corner states develop. The  size of the system is $5\times 5$ unit cells, $\gamma/\kappa=0.17$.
\textbf{c} Corner charge, edge dipole polarizations and bulk quadrupole polarization depending on the flux $\pi$. The calculation has been performed for  $N=30$ unit cells, and $\gamma/\kappa=0.17$. In order to fix the sign of the quadrupole momentum the diagonal energies at the diagonal sites of each unit cells were set to $\pm\delta=\pm 10^{-4}\kappa$~Ref.~\cite{Benalcazar2017b}
}
\label{fig:S5}
\end{figure}

The schematics of the system under study is shown in Fig.~\ref{fig:S5}~\textbf{a}. We assume that the couplings $\gamma$ and $\kappa$ are purely real, while some of the couplings in the lattice are complex and have an extra hopping phase: $\gamma\,e^{i\phi}$ and $\kappa\,e^{i\phi}$. Note that this geometry ensures $\phi$ flux through each square plaquette of the lattice. The momentum space Hamiltonian of the periodic structure reads
\begin{equation}
H(k_{x},k_{y})=\begin{pmatrix}
0&\gamma+\kappa\e^{i k_{y}}&0&\gamma+\kappa\e^{i k_{x}}\\
\gamma+\kappa\e^{-i k_{y}}&0& e^{i\phi}(\gamma+\kappa \e^{i k_{x}})&0\\
0& e^{-i\phi}(\gamma+\kappa \e^{-i k_{x}})&0&\gamma+\kappa \e^{-i k_{y}}\\
\gamma+\kappa e ^{-i k_{x}} &0&\gamma +\kappa \e^{i k_{y}}&0
\end{pmatrix}\:,
\end{equation}
where $k_{x}$ and $k_{y}$ are the Cartesian projections of the Bloch wave vector. Similarly to the article main text, we choose the ratio of the hopping amplitudes $\gamma/\kappa=0.17$.

Zero flux case corresponds to the tight-binding system with all positive couplings which can be viewed as a 2D analogue of the Su-Schrieffer-Heeger (SSH) model. The spectrum of this system has no bandgap at zero energy and for that reason does not feature any robust corner states. However, when some of the couplings are made complex such that nonzero gauge flux appears, the bandgap opens and the degenerate corner states emerge, as shown in Fig.~\ref{fig:S5}~\textbf{b}. The width of the bandgap grows as the gauge flux is further increased up to $\pi$, thus enhancing the robustness of the corner states. The system also hosts one-dimensional edge states at $\delta\nu\approx \pm \kappa$, that are localized at the sides of the square and weakly depend on the flux.

\linespread{1.0}

Even more instructive picture is provided by the evolution of dipole and quadrupole polarizations in the system. The dependence of the polarizations on the flux is shown in Fig.~\ref{fig:S5}~\textbf{c}. The corner charge $Q_{\rm corner}$ has been calculated for a large 2D system with open boundary conditions at all 4 sides.  It was found to be equal to $1/2$ independent of the flux.  The dipole polarizations $p_{x}=p_{y}$ were evaluated following the procedure in Ref.~\cite{Benalcazar2017b}. This calculation involves   opening the boundaries along $x$, using periodic boundary along $y$ direction, and evaluating the dipole polarizations from the averaged positions of the Wannier centers. Next, the bulk quadrupole momentum has been evaluated from the equation\:.
\begin{equation}
Q_{\rm corner}=p_{x}+p_{y}-q_{xy}\:. \label{eq:corner}
\end{equation}
In agreement with the general theory, Fig.~\ref{fig:S5}~\textbf{c} demonstrates, that for the quadrupolar insulator, where $\phi=\pi$, the corner charge is contributed by both quantized dipole polarizations and bulk quadrupole momentum.  When the flux is unequal to $\pi$ or $0$, the value of $q_{xy}$ in general is not quantized~\cite{Benalcazar2017b}. When the flux decreases the bulk quadrupole momentum vanishes and becomes zero in the 2D SSH case, the edge polarizations decrease as well, but the corner charge stays the same.  Our calculation indicates that the scale on which the quadrupole momentum changes  from $0$ to $1/2$ is determined by the small on-site potential $\pm \delta=\pm 10^{-4}\kappa$. The energies $\delta$, $-\delta$, $\delta$, $-\delta$ in the clockwise order are added to the sites of the unit cell in order to fix the sign of the corner charges and the quadrupole momentum ~\cite{Benalcazar2017b}.

The case of zero flux is degenerate and requires special care, since the 2D SSH system is not an insulator, i.e. it lacks a band gap. Our findings of $p_{x}=p_{y}=1/4$, $Q_{\rm corner}=1/2$ and $q_{xy}=0$ can be also understood from the results for the anisotropic 2D SSH model with $\kappa_{x}\ne \kappa_{y}$ that has a band gap and well-defined polarizations quantized by chiral and inversion symmetries. Namely, in the anisotropic 2D SSH case with $\kappa_{x}>\kappa_{y}$ one has $p_{x}=1/2$ and $p_{y}=0$, while for $\kappa_{y}>\kappa_{x}$  the polarizations are $p_{y}=1/2$ and $p_{x}=0$~\cite{Benalcazar2017b}. The isotropic model with  flux tending to zero is an intermediate case where the edge polarizations have to be equal. Thus, averaging the results for $\kappa_{x}>\kappa_{y}$  and $\kappa_{x}<\kappa_{y}$  we obtain $p_{x}=p_{y}=1/4=Q_{\rm corner}/2$,\quad   $q_{xy}=0$, in agreement with the rigorous calculation in Fig.~\ref{fig:S5}~\textbf{c}.

The essence of the topological transition as a function of the gauge flux $\phi$ can be understood with the following symmetry arguments.  The quadrupole topological phase is protected by the reflection symmetries along the $x$ and $y$ axes, denoted by $M_x$ and $M_y$ respectively.  In order to have protected topological phase and quantized quadrupole moments, one needs to satisfy the following necessary conditions (see Ref.~\cite{Benalcazar2017b} for derivation):
\begin{enumerate}
\item
$[M_x, H]=[M_y, H]=0$, i.e., the system Hamiltonian $H$ commutes with the two reflection symmetries.
\item
$[M_x, M_y] \neq 0$, i.e., the two reflection symmetries do not commute.  This condition ensures that the edge states are gapped, which is equivalent to gapped ``Wannier bands" (see Ref.~\cite{Benalcazar2017b} for the definition and discussion).
\end{enumerate}
It is straightforward to see that condition 1 is only satisfied at $\phi=0, \pi$. Therefore, for any $\phi$ in between, the quadrupole moment $q_{xy}$ is not quantized and hence not topological protected, as indicated in Fig.~\ref{fig:S5}~\textbf{c}.    On the other hand,  at the $\phi=0$ point, condition 2 is not satisfied, while it is satisfied at the $\phi=\pi$ point due to the introduction of gauge flux.   Therefore, the $\phi=0$ point generically has gapless edge states, or equivalently gapless Wannier bands (in the anisotropic case $\kappa_x \neq \kappa_y$), leading to unprotected corner states.  The isotropic case $\kappa_x = \kappa_y$ is special since the bulk band is gapless as stated above, so the Wannier band is not well defined.  In this case, the fact that the corner states are unprotected can be attributed to the absence of the bulk band gap.

\bibliographystyle{NatureMag}
\bibliography{Quad_Topo_Biblio}